\documentclass[12pt]{article}
\usepackage{graphicx}
\usepackage{epsfig}
\usepackage{amsmath}
\usepackage{amssymb}
\usepackage{amsthm}
\usepackage{graphics}

\newcommand{\beq}{\begin{equation}}
\newcommand{\beqn}{\begin{eqnarray}}
\newcommand{\eeq}{\end{equation}}
\newcommand{\eeqn}{\end{eqnarray}}

\newcommand{\lp}{\left}
\newcommand{\rp}{\right}
\newcommand{\la}{\langle}
\newcommand{\ra}{\rangle}

\newcommand{\eff}{{\rm eff}}

\newcommand{\tmn}{\la T_{\mu\nu}\ra}

\newcommand{\ricc}{{\cal R}}
\newcommand{\sign}{{\rm\footnotesize sign}}
\def\beq{\begin{equation}}
\def\beqn{\begin{eqnarray}}
\def\eeq{\end{equation}}
\def\eeqn{\end{eqnarray}}
\def\lp{\left}
\def\rp{\right}
\def\la{\langle}
\def\ra{\rangle}
\def\eff{{\rm eff}}

\def\mpl{M_{\rm Pl}}

\def\ricc{{\cal R}}

\def\lsim{\mathrel{\rlap{\lower3pt\hbox{\hskip0pt$\sim$}}
     \raise1pt\hbox{$<$}}}         
\def\gsim{\mathrel{\rlap{\lower4pt\hbox{\hskip1pt$\sim$}}
     \raise1pt\hbox{$>$}}}         

\newcommand{\R}{\mathcal{R}}
\newcommand{\Tdw}{\hat{T}^{\,\mathrm{DW}}}

\newcommand{\GN}{G_{\mathrm{N}}}
\newcommand{\Z}{\mathbb{Z}}

\numberwithin{equation}{section}

\input epsf%
\begin{document}


\centerline{\Large \bf  Dressed Domain Walls}~\\ %
\centerline{\Large \bf  and Holography}

\vspace{1cm}

\centerline{\large Luca Grisa\footnote{luca.grisa@physics.nyu.edu}
and Oriol Pujol{\`a}s\footnote{pujolas@ccpp.nyu.edu}}

\vspace{5mm}

\centerline{\emph{Center for Cosmology and Particle Physics}}
\centerline{\emph{Department of Physics, New York University} }
\centerline{\emph{New York, NY 10003, USA} }

\vspace{1cm}

\begin{abstract}

The cutoff version of the AdS/CFT correspondence states that the
Randall Sundrum scenario is dual to a Conformal Field Theory (CFT)
coupled to gravity in four dimensions.
The gravitational field produced by relativistic Domain Walls can
be exactly solved in both sides of the correspondence, and thus
provides one further check of it.
We show in the two sides that for the most symmetric case, the
wall motion does not lead to particle production of the CFT
fields.
Still, there are nontrivial effects.
Due to the trace anomaly, the CFT effectively renormalizes the
Domain Wall tension.
On the five dimensional side, the wall is a codimension 2 brane
localized on the Randall-Sundrum brane, which pulls the wall in a
uniform acceleration.
This is perceived from the brane as a Domain Wall with a tension
slightly larger than its bare value.
In both cases, the deviation from General Relativity
appears at nonlinear level in the source, and the leading
corrections match to the numerical factors.
%
%

\end{abstract}

\newpage

\tableofcontents

\section{Introduction}
\label{sec:intro}

The AdS/CFT correspondence provides a powerful tool to investigate
the dynamics of gauge theories. In its original form, the
correspondence relates a four dimensional ${\cal N}=4$ $SU(N)$
Super Yang-Mills (SYM) at large $N$ and strong  coupling without
gravity to a 5D theory of gravity in Anti de Sitter (AdS) space
\cite{malda,witten,klebanov} (see \cite{magoo} for a review). In
this note, we will be interested in an extension of this
holographic duality where gravity is also included in the 4D side.
This corresponds to chopping off the AdS bulk by the presence of a
(UV) brane \cite{verlinde,gubser,ahpr} as in the Randall Sundrum
(RS) model \cite{rs2}. Hence, the gravitational dynamics of matter
localized on a RS brane is dual to a 4D setup where the quantum
effects of the Conformal Field Theory (CFT)  are taken into
account. This is a particularly convenient way for learning about
the quantum effects of field theories in nontrivial gravitational
backgrounds, and has already been exploited in a variety of
situations ranging from cosmology
\cite{gubser}--\nocite{hhr,verlindeRadiation,se,no,shiromizu1,shiromizu2}\cite{takahiroFRW}
to Black Hole physics
\cite{chr}--\nocite{gs,ehm1,ehm2,gt,DuffLiu,takahiroBH,efk,egbk,germani1,balbinotFabbri,germani2,frw,FabbriProcopio,fabbriolmo2}\cite{takahiroFloating}.

Here, we shall revisit the quantum effects present on the
gravitational background produced by (relativistic) Domain Walls,
from the new perspective offered by the correspondence. The
interest for this case is that the DW represents a physical
situation with a moving mirror and it may give rise to particle
creation
\cite{BD}--\nocite{fullingDavies1,fullingDavies2,cd,frolovSerebriany,cw1}\cite{cw2}.
Given the similarity between this and Hawking radiation, this
might shed some light on the ongoing debate concerning black holes
in the Randall Sundrum scenario.
As argued in \cite{takahiroBH,efk}, the expectation is that the
large number of fields of the CFT enhances the evaporation rate
and that there should be no static solutions.
However, given that the CFT is strongly coupled, it is not
entirely clear that the number of states to which the BH can
radiate is of order $N^2$ \cite{frw} (see also
\cite{FabbriProcopio,takahiroFloating}).
On the other hand, the DW case is more tractable. In particular,
the amount of radiation produced by the wall can be explicitly
found.
As a first step, in this note we will restrict our attention to
the cases without particle production.
As we will see, even in this case the situation is not entirely
trivial.
Furthermore, for the most symmetric configurations, the problem
can be exactly solved in both sides of the correspondence and,
hence, provides one more test of its validity.\\

In General Relativity (GR), the spacetime generated by a DW with a
maximally symmetric worldvolume is given by the so-called
Vilenkin-Ipser-Sikivie (VIS) spacetime \cite{vilenkinDW,is}. In
this solution, the DW inflates at a rate $H_0$ determined by its
tension $\sigma$ as $H_0=\sigma/4\mpl^2$ ($\mpl$ is the Planck
mass) and there is a Rindler horizon. This represents a repulsive
gravitational field, since test particles are repelled from the
wall with an acceleration set
by $H_0$. \\

In the holographic dual of the Randall Sundrum model, the DW
gravitates accordingly with GR but the quantum effects from the
CFT (which couples to the DW only through gravity) \emph{and their
backreaction} are also included.
Generically, the CFT can produce two types of effects. The first
is particle creation. As we will see, as long as the DW
worldvolume is maximally symmetric and the spacetime has a
horizon, then no CFT modes can be produced. This is related to the
classic result for moving mirrors, which do not create particles
of conformal fields when the mirror moves with uniform
acceleration.

The other kind of effect is to modify the way how the metric
responds to the source, essentially due to the trace anomaly. In
our case, the DW tension $\sigma$ is effectively renormalized by
an amount of order $N^2\,H_0^3$.
As a result, one can obtain self-consistent solutions representing
the spacetime produced by the DW \emph{dressed} with the CFT
corrections. What one finds is that the Hubble rate on the wall
(and the gravitational repulsion that it produces) is larger than
without the CFT.

One property of the self-consistent solutions is that they cease
to exist for tension larger than a certain critical value
$\sigma_c^{(4)} \sim \mpl^3 /N$. This happens precisely when the
curvature scale of the DWs becomes comparable to the cutoff, which
is of order $\mpl/N$. Hence, for $\sigma\gtrsim\sigma_c^{(4)}$ the
theory breaks down and the details of the UV completion are
important. If one views the Randall Sundrum model as the
completion, this corresponds to the transition to the regime where
gravity behaves as in 5D.\\

%

The 5D dual of the dressed walls are DWs localized on a RS brane
\cite{rs2}. In this context, a DW is a codimension 2 brane, and
exact solutions can also be easily found. The DW produces a
deficit angle solely determined by its tension, $\sigma$. On the
other hand, the RS brane `pulls' the codimension 2 brane in an
accelerated motion. As a result, and in contrast with what happens
in isolation, codimension 2 branes `attached' to a codimension 1
brane effectively generate a repulsive gravitational field.
Hence, from the point of view of a four dimensional observer on
the brane, the wall \emph{appears} as an ordinary Domain Wall. As
we will see, the induced metric on the brane is of the VIS form,
and only differs from GR in how the Hubble rate on the wall $H_0$
relates to $\sigma$. Furthermore, the leading order deviation from
the GR result exactly matches the one we find in the CFT side,
providing one further check of the correspondence.

The fact that the DWs start behaving according to 5D gravity for
tensions larger than the critical value $\sigma_c^{(4)}$ is
slightly more involved to see, but it can be summarized as
follows.
In the thin wall approximation, the worldvolume curvature scale
$H_0$ diverges when the tension approaches $\sigma_c^{(5)}\equiv
2\pi M^3$, where $M$ is the Planck mass in the bulk. This is the
codimension 2 notion of critical tension (the deficit angle that
it produces is $2\pi$), and it is of the same order as
$\sigma_c^{(4)}$. Of course, the divergence in $H_0$ only means
that sooner or later one has to resolve the wall thickness, $d$.
In doing so, one realizes that $H_0$ becomes finite even for
`supermassive' walls, that is, with $\sigma > \sigma_c^{(5)}$. If
the thickness is small enough ($d\ll \ell$) then the AdS curvature
becomes irrelevant and, in the neighborhood of the DW, one of the
two transverse directions is effectively compactified. Since the
compactification radius is of order $d$, the effective Planck mass
felt by the DW is of order $d M^3$, which is much smaller than the
4D Planck mass, $\ell M^3$. The outcome is this: for supermassive
walls, $H_0$ is enhanced with respect to GR by a factor $\ell /
d$.
As it was described in \cite{dgpr}, codimension 2 branes display
precisely the same kind of behaviour when the tension becomes
supercritical (they compactify one of the transverse directions
and inflate as a DW).
Hence, the gravitational field of localized DWs turns five
dimensional when the tension becomes roughly of order
$\sigma_c^{(5)}\simeq\sigma_c^{(4)}$, as expected from the CFT
side.\\

This paper is organized as follows. In Section \ref{sec:GR} we
review the gravitational effects of Domain Walls in GR. In Section
\ref{sec:cft} we work out the self-consistent solution
incorporating the CFT corrections. We discuss the 5D dual in
Section \ref{sec:RS}, finding the solutions for DWs localized on
the brane in Sec.~\ref{sec:sol} and \ref{sec:structure}. We also
show in Sec.~\ref{sec:RSthickness} how resolving the thickness
allows to have supermassive DW solutions.  We compare the results
from the two sides of the correspondence in \ref{sec:comparison},
and we will conclude in Section~\ref{sec:conclusions}.

\section{Domain Walls in GR}
\label{sec:GR}

In this Section, we shall briefly review some of the properties of
gravitating Domain Walls in General Relativity (GR). We shall
start by the thin wall description, assuming that the wall
thickness is much smaller than any other scale of the problem and
treating the wall as an idealized distributional source. We
comment on the case with finite thickness in Sec.~\ref{sec:thick}.

In the thin wall approximation, the stress tensor of a Nambu-Goto
Domain Wall takes the form
\begin{equation}\label{tmunuThin} %
T^{DW}_{\mu\nu}= \sigma \, \delta(\xi) \; {\rm
diag}\lp(1,-1,-1,0\rp)_{\mu\nu}
\end{equation}
where $\sigma$ is the `tension' or surface energy density,
$\delta(\xi)$ is the Dirac $\delta-$function and $\xi$ represents
a `proper' coordinate normal to the wall, which one can always
find by going to Gaussian Normal coordinates around the
hypersurface defined by the DW worldvolume itself.

We shall concentrate on the  DW solutions with a maximally
symmetric worldvolume and with $\mathbb{Z}_2$ symmetry across the
wall.
One can easily see that this implies that
%
the metric must admit a 3D maximally symmetric slicing,
\emph{i.e.}, it is of the form
\begin{equation}\label{induced}%
ds_4^2=d\xi^2+R^{2}(\xi)\;ds_{\kappa}^2\,. 
\end{equation}
where $ds_{\kappa}^2$ is the line element of a 3D Minkowski
($\kappa=0$), de Sitter ($\kappa=1$) or Anti de Sitter
($\kappa=-1$) spacetime of unit radius. From \eqref{induced}, the
curvature scale of the DW worldvolume is determined by $R(0)$.
Thus, we shall introduce the Hubble rate on the DW as\footnote{In
our notation, $H_0^2<0$ means that the worldvolume is AdS${}_3$.}
$$
H_0^2 \equiv {\kappa\over R^2(0)}~.
$$

We shall consider the case when a cosmological constant
$\Lambda_4$ is also  present. With a metric of the form
\eqref{induced}, the Einstein equations become
\begin{align}
3\mpl^2\;\frac{\kappa-R'^{\,2}}{R^{\,2}}&=\Lambda_4\,, \label{GR.a} \\
\mpl^2\;\lp(\frac{\kappa-R'^{\,2}}{R^{\,2}}-2\frac{R''}{R}\rp)&=
\sigma\delta(\xi)+\Lambda_4\, . \label{GR.b}
\end{align}
Integrating \eqref{GR.b} around $\xi=0$ leads to the junction
condition on the DW
\begin{equation}\label{junctionGR} %
K_0 = {\sigma \over 4 \mpl^2} ~, 
\end{equation} %
where we have introduced
\begin{equation}\label{K0} %
K_0 \equiv -{1\over2}{\Delta R'(0) \over R(0)} ~, 
\end{equation} %
and $\Delta X$ denotes $X(\xi=0^+)-X(\xi=0^-)$. With this, $K_0$
is proportional to the (jump in the) extrinsic curvature of the
wall as embedded in the 4D geometry \eqref{induced}. As we
illustrate below, physically $K_0$ sets the scale of the
accelerated motion of the DW or, equivalently, the scale for the
gravitational repulsion that the DW exerts on test particles.
Equation \eqref{junctionGR} thus summarizes the gravitational
effect of a DW in GR. In the next sections, we will find
deviations from this relation.

On the other hand, the curvature scale on the DW worldvolume
depends on the properties of the ambient space. The curvature
scale away from the wall is
$$
H_4^2 =  
 {\Lambda_4 \over 3 \mpl^2}~.
$$
Hence, Eq. \eqref{GR.a} evaluated at $\xi=0$ can be rewritten as
\begin{equation}\label{GaussCodacci} %
H_0^2=K_0^2+H_4^2~. 
\end{equation}
This is the familiar Gauss equation for the embedding of the DW in
the ambient 4D space. It follows that in de Sitter and in flat
space, the walls always inflate, while in AdS${}_4$ only DWs with
tension
larger than $4 \mpl \sqrt{|\Lambda_4|/3}$ do so. %
For inflating DWs ($\kappa=1$), the `warp factor' that solves
\eqref{GR.a} and \eqref{GR.b} is
\begin{equation}\label{warpGeneral}
R(\xi)=H_4^{-1}\sin \lp[\sin^{-1}(H_4/H_0)  -H_4|\xi|\,\rp]~.
\end{equation}
which is valid for any $\Lambda_4$ by analytic continuation. As
one can see, in these cases there is a horizon ($R(\xi)$ vanishes)
at a finite proper distance from the wall.

\begin{figure}[!tb]
\begin{center}
  \includegraphics[height=6cm]{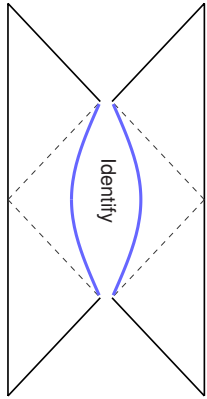}%
\end{center}
\caption{Conformal diagram of the Vilenkin Ipser Sikivie
spacetime \cite{Gibbons:1993in}, the metric produced by a Domain Wall in otherwise flat
space. The (inflating) DW corresponds to the hyperboloid
represented by the blue line. The Rindler horizon (dashed line) is
at $\xi=\pm 1/H_0$, and separates the Rindler and Milne regions. }
\label{fig:VIS} %
\end{figure}

To gain some intuition, let us briefly look at the full form of
the metric for the $\Lambda_4=0$ case. The warp factor is
\begin{equation} \label{vis}
  R(\xi)=H_0^{-1}-|\xi|~,
\end{equation}
giving what is usually referred to as the Vilenkin-Ipser-Sikivie
spacetime \cite{vilenkinDW,is}. The conformal diagram for this
space is shown in Fig. \ref{fig:VIS}. It can be viewed as two
copies of the interior of the hyperboloid defined by the DW, glued
together at the DW position. The coordinate $\xi$ covers part of
the Rindler patch of Minkowski, and the Rindler (or
`acceleration') horizon is at $\xi=\pm 1/H_0$.

An inertial observer in this space sees the wall moving away at a
constant acceleration set by $K_0$ (which coincides with $H_0$ in
this case), effectively feeling a \emph{repulsive} gravitational
field. In this sense, the DWs unambiguously produce a nontrivial
gravitational effect even though the curvature tensor is
identically zero (away from the DW) and so there are no tidal
forces. For general $\Lambda_4$, the scale of the
acceleration/repulsion is set by $K_0$ but the curvature of the DW
worldvolume (and the presence of the horizon) depends on
$\Lambda_4$.

\subsection{Finite thickness walls}
\label{sec:thick}

In the above discussion, the DW is treated as infinitesimally thin
and its stress tensor as a distribution. This is not a good
approximation when the curvature radius of the worldvolume is
comparable to the thickness, $d$, of the wall. For our purposes,
the most relevant situation when this happens is when the
curvature scale provided by the gravity of the wall $K_0$ becomes
comparable to $1/d$, that is when
\begin{equation}
  \sigma \gtrsim {\mpl^2 \over d}~.
\end{equation}
For $\Lambda_4=0$, this corresponds to a regime of
\emph{topological inflation} \cite{vilenkinTI}, since the Hubble
radius on the wall is smaller than its actual thickness, and so
the DW is also inflating in the transverse direction.


We shall now show that when we take into account a nonzero $d$,
then the pressure in the direction transverse to the (gravitating)
DW must be nonzero, and one should replace \eqref{tmunuThin} by
\begin{equation}\label{tmunu} %
T_\mu^\nu= {\rm diag}\lp(-\rho,-\rho,-\rho,P\rp)~.
\end{equation}
To make this compatible with the same symmetries, $P$ and $\rho$
must depend on $\xi$ only. On the other hand, the local
conservation equation on a space of the form \eqref{induced} leads
to
\begin{equation}\label{cons} %
P'+3{R'\over R}(P+\rho)=0~. 
\end{equation}
We do not need to specify the actual microscopic model for the
Domain Wall, as our conclusions will  be model independent. The
only thing we need to know is that the stress tensor of the wall
obeys \eqref{cons}, and $\rho(\xi)$ is peaked at $\xi=0$, decaying
exponentially fast to zero for $|\xi| \gtrsim d$, and that
$R(\xi)$ is similarly a smoothed-out version of
\eqref{warpGeneral}. These assumptions and \eqref{cons} suffice to
show that $P(\xi)\neq0$ in the DW core. Note that this is
independent of whether the DW is inflating or not, rather it
requires only a nonzero extrinsic curvature, $R'/R$. Hence, the
presence of $P\neq0$ can be thought to be necessary to support the
accelerated motion of the DW.

In this respect, note that the thin wall approximation advocated
before is slightly inaccurate. Indeed, \eqref{tmunuThin} assumes
$P=0$ even on the DW core, so that \eqref{cons} leads to $R'(\xi)
\delta(\xi)=0$. While this is not very well defined as a
distribution ($R'$ is discontinuous at $\xi=0$), it is not clear
that this should be treated as $0$ even in a distributional sense.
Instead, a pressure of the form $\propto 1-\varepsilon^2(\xi)$,
where $\varepsilon(\xi)=\sign(\xi)$ for $\xi\neq0$ and
$\varepsilon(0)=0$, is a more accurate representation compatible
with \eqref{cons}. While it remains true that this distribution
vanishes when integrated with any smooth test function, it
captures the fact that for gravitating thick DW solutions, it is
never strictly zero at the core of the wall
\cite{widrow}.\footnote{In contrast, the situation for cosmic
strings is qualitatively different. For sub-critical (\emph{i.e.},
such that the deficit angle is less than $2\pi$) BPS strings, the
pressure in the transverse directions remains zero in the core
even including gravity. }

Actually, incorporating $P$ is also quite convenient because the
(appropriately modified version of the) constraint \eqref{GR.a} is
a first order equation analogous to the Friedmann equation in
cosmology, and it suffices to find the solutions to Einstein
equations, once \eqref{cons} is enforced. Indeed, \eqref{GR.b} is
a consequence of \eqref{cons} and \eqref{GR.a} as long as $R(\xi)$
is not constant.


\section{Quantum effects from the CFT}
\label{sec:cft}

In this Section we show how the quantum effects from the CFT
modify the gravitational field of a Domain Wall discussed in the
previous Section. We are assuming that the DW is a source external
to the CFT which only interacts with it through gravity. The field
produced by the DW can be simply computed from the graviton
propagator \emph{dressed} with the radiative corrections from the
CFT. This is the procedure followed in \cite{DuffLiu} for a point
particle, where a precise agreement with the brane world result of
\cite{gt} was found.

In the DW case, this leads immediately to the conclusion that the
CFT does not modify --at all-- the field produced by DWs (\emph{to
linear order} in the source). This follows from the K\"allen
Lehmann decomposition of the graviton propagator, when the CFT
radiative corrections are included. The propagator has a massless
graviton pole and a branch cut, which are dual to the massless
graviton and the continuum of Kaluza Klein modes in the RS model.
The key point is that the tensor structure of the propagator
associated with the branch cut is granted to be that of a massive
graviton. This couples to conserved sources as
$h^{massive}_{\mu\nu}\sim
T_{\mu\nu}-(T_\rho^\rho/3)\eta_{\mu\nu}$, and as it was observed
in \cite{dgpr}, this leads to a pure gauge metric perturbation.
Hence, only the massless pole contributes to linear order in the
graviton exchange with a Domain Wall and no correction is found
even though the propagator differs from that of pure GR. In the
Appendix, we illustrate it by performing a more explicit
computation at 1-loop.

In the  rest of this Section we shall show that, despite the
previous argument, the CFT \emph{does} induce a correction to the
DW gravitational field that is nonlinear in the tension $\sigma$.
The reason is that in the 4D side of the correspondence, the
backreaction from the quantum effects of the CFT is also
incorporated. This implies that the 4D metric obeys the
semiclassical Einstein equations
\begin{equation}\label{semiclassical}
  \mpl^2 \, G_{\mu\nu}= T_{\mu\nu}^{DW} + \tmn^{CFT}~,
\end{equation}
$\tmn^{CFT}$ is the vacuum expectation value of the stress tensor
of the CFT. The solutions of \eqref{semiclassical} differ from GR
because $\tmn^{CFT}\neq0$ in the VIS spacetime described in
Sec.~\ref{sec:GR}, essentially because the trace anomaly is
nonzero in this background (see below). However, one can easily
find \emph{self-consistent} solutions of \eqref{semiclassical}.
One takes an `ansatz' for the metric of the form \eqref{induced},
with $R(\xi)$ to be determined. Since for a maximally symmetric DW
worldvolume, the form of $\tmn^{CFT}$ can be determined without
explicit reference to the actual form of $R(\xi)$, then
\eqref{semiclassical} fixes the function $R(\xi)$ and hence the
self-consistent solution.

Let us now see how to determine the energy-momentum tensor
$\tmn^{\mathrm{CFT}}$. We can always separate it as
\begin{equation}\label{tmncft}
  \tmn^{CFT}= \tmn^{(0)}+ T_{\mu\nu}^{\mathcal{A}}
\end{equation}
where $T_{\mu\nu}^{\mathcal{A}}$ is the anomalous contribution
while $\tmn^{(0)}$ is tracefree and depends on the choice of
vacuum.
The nonzero trace ${T_{\mu}^{\mu}}^{\mathcal{A}}$ arises from the
anomalous breaking of conformal invariance by the quantum effects
and may be non-zero in curved spacetimes.


\subsection{Absence of particle production}

\label{sec:absence}

The state dependent contribution $\tmn^{(0)}$ in \eqref{tmncft}
encodes the effects from particle creation. As we shall see,
whenever there is a horizon and the wall is maximally symmetric,
then $\tmn^{(0)}=0$ and the wall does not radiate quanta of the
CFT.

We will assume as before that the metric solving
\eqref{semiclassical} allows a maximally symmetric slicing and, as
such, it is of the form \eqref{induced}. In this case, the
state-dependent piece $\tmn^{(0)}$ can be completely fixed as
follows. On the one hand, conformal invariance demands it to be
traceless. On the other, imposing that it has the symmetries of
the metric \eqref{induced} leads to
\begin{equation}
    \la T_\mu^{\,\nu}\ra^{(0)}=p(\xi)\, {\rm diag} \big(-\frac{1}{3},-\frac{1}{3},-\frac{1}{3},1\big)^\nu_\mu\,,
    \label{statedep.T}
\end{equation}
where the function $p$ depends on the DW transverse direction
$\xi$ only. This kind of stress tensor (for any nontrivial
dependence on the `Rindler' coordinate $\xi$) implies the presence
of fluxes of energy-momentum away from the DW, which is
interpreted as arising from the particle creation by the moving
mirror \cite{BD}.

The  function $p(\xi)$ is further determined from the conservation
of the stress-energy tensor, which leads to $p'/p+4R'/R=0$, where
$R(\xi)$ is the warp factor as defined in \eqref{induced}. Thus,
$p(\xi)$ must be of the form $p_0/R^4(\xi)$, for some
constant\footnote{Hence, based entirely on the symmetries, one can
fix the form of $\tmn^{CFT}$ up to a constant. As will become
apparent in Sec.~\ref{sec:RS} this argument is the holographic
dual of the Birkhoff theorem.} $p_0$. If the effects from the CFT
do not change the asymptotic structure of the spacetime, then
there is a Milne region (see Fig. \ref{fig:VIS}) and hence a
horizon%
\footnote{From now on, we restrict ourselves to the case when
there is a horizon, that is $\kappa=0,1$. This excludes DWs with
small enough tension in Anti de Sitter, with an AdS${}_3$
worldvolume. This case will be considered elsewhere. }.
That is, $R(\xi)$ vanishes somewhere. Were $p_0$ nonzero,
$\tmn^{(0)}$ would be singular on the horizon. Thus, the only
vacuum state of the CFT that is regular on the horizon has
$p_0=0$.

Given that the 4D side of the correspondence is supposed to
include the backreaction from the CFT, we should take it into
account before reaching a final conclusion. In particular, we
should make sure that the backreaction is not going to change the
assumption that there is a horizon.
Ignoring the anomalous contribution for simplicity, this reduces
to finding the possible solutions of the `Friedmann' equation
$3\mpl^2 (1-R'^{\,2})/R^{\,2}=-p_0/R^4$.
One can easily show that if $p_0$ were positive, then the horizon
is replaced by a naked singularity. Thus, the quantum state would
be indeed singular and we should discard it as unphysical. On the
other hand, if $p_0$ were negative then there would be regular
`bouncing' solutions, with flat asymptotics ($R'(\xi)\to \pm1$ for
large $\xi$) and no horizons. However, since $p_0$ would be (at
most) of order $N^2$, the curvature scale would be (at least) of
order of the cutoff around the bounce. Hence, these solutions
cannot be trusted and we should disregard them as well. This is
confirmed by the results of Sec.~\ref{sec:RS}, where no analogous
solution is found.

%

Thus, we conclude that indeed in the regular vacuum state there is
no particle creation, {\em i.e.}, $p_0=0$.\footnote{Let us add
that physical solutions with $p_0<0$ (and no horizon) may arise if
the direction transverse to the DW is compactified. In that case,
a Casimir energy may appear. If the compactification length is
small enough, then $p_0$ can exceed the $\sim N^2$ estimate. Then,
the curvature scale at the bounce can be below the cutoff and the
solution can be trusted. As we shall see in Sec.~\ref{sec:RS}, the
duals of these solutions in 5D are AdS${}_5$ bubbles of nothing
\cite{WittenBON}--\nocite{afhs,birmingham,balaRoss}\cite{balaKlausSimon}.
In this paper, though, we shall only consider the case when $\xi$
is noncompact, as in the solutions described in
Sec.~\ref{sec:GR}.}
This result is confirmed by the explicit
1-loop computations \cite{cd}--\nocite{frolovSerebriany,cw1}\cite{cw2},\cite{pt} for scalar fields%
\footnote{These calculations also show that the vanishing of
$\tmn^{(0)}$ is independent of what kind of boundary conditions
satisfied by the CFT fields at the DW location.
Indeed, $\tmn^{(0)}$ is found to vanish even introducing a
coupling of the scalar field to the DW of the form $m_0
\delta(\xi) \phi^2$ \cite{pt}. The only effect of this at 1-loop
is to renormalize the DW tension \cite{ps}.
Let us also emphasize that for non-conformal fields $\tmn^{(0)}$
does not vanish (and is regular on the horizon)
\cite{cd}--\nocite{frolovSerebriany,cw1}\cite{cw2},\cite{xavi},\cite{pt}.},
and is also expected from the analogy with moving mirrors
\cite{BD}. As is well known, they only radiate to conformal fields
if the motion is \emph{not} uniformly accelerated, that is, if the
worldvolume of the mirror is \emph{not} maximally symmetric.
Let us emphasize that this result should be viewed as a statement
about the incompatibility of having particle production in the CFT
with the amount of symmetries assumed. As such, it should remain
true to all orders in the loop expansion.
This will be confirmed by the computation in the dual setup of
Sec.~\ref{sec:RS}.

\subsection{The anomalous contribution}

The previous discussion implies that (for maximally symmetric
worldvolume and when the Rinlder horizon is present) the only
contribution to $\tmn^\mathrm{CFT}$ may come from the conformal
anomaly. Let us now work out this contribution. It is known that
in 4D at first loop the anomalous trace is given by
\begin{equation}
{T^\mu_\mu}^{\,\cal A} =
    -a\,E_{(4)}+c\,I_{(4)}+d\,\nabla^2\ricc\,,
    \label{conf.anom}
\end{equation}
where $E_{(4)}$ is the Euler density
\begin{eqnarray}
E_{(4)}&=&\R_{\mu\nu\rho\sigma}\R^{\mu\nu\rho\sigma}-4\R_{\mu\nu}\R^{\mu\nu}+\R^2\,,
\end{eqnarray}
and $I_{(4)}=C_{\mu\nu\rho\sigma}C^{\mu\nu\rho\sigma}$ with
$C_{\mu\nu\rho\sigma}$ the Weyl tensor.

The parameters $a$, $c$, and $d$ depend on the field content of
the theory and are found to be
\begin{eqnarray}
a&=&\frac{1}{360\,(4\pi)^2}(N_0+\frac{11}{2}N_{1/2}+62N_1)\,,\\
c&=&\frac{1}{120\,(4\pi)^2}(N_0+3N_{1/2}+12N_1)\,,\\
d&=&\frac{1}{180\,(4\pi)^2}(N_0+3N_{1/2}-18N_1)\,,
\end{eqnarray}
where $N_0$, $N_{1/2}$, and $N_1$ are respectively the number of
scalars, Majorana fermions and vectors of the field theory. In the
particular case of RS, the field content is that of
$\mathcal{N}=4$ Super Yang-Mills Theory, therefore $N_0=6N^2$,
$N_{1/2}=4N^2$ and $N_1=N^2$, with $N$ the rank of the gauge
group. With this field content, one finds
\begin{equation}
a=c=\frac{N^2}{64\pi^2}\,,\qquad d=0\,.
\end{equation}
In the literature, a finite counterterm of the form $\sqrt{-g}\,
R^2$ is often added \cite{starobinsky,vilenkinAnomaly,hhr}, and it
has the effect of shifting $d$ away from $0$. This term explicitly
breaks conformal invariance, and we assume it is not present.


Once ${T^\mu_\mu}^{\,\cal A}$ is known, the symmetries of the
problem fix the full form of $T_{\mu\nu}^{\cal A}$ as
${T_{\mu}^\nu}^{\,\cal A} ={\rm diag }(-\rho_{\cal A},-\rho_{\cal
A},-\rho_{\cal A},p_{\cal A})$ with $p_{\cal A}$ and $\rho_{\cal
A}$ obeying the conservation equation ${p_{\cal A}}'+3(p_{\cal
A}+\rho_{\cal A}) R'/R=0$ and Eq. \eqref{conf.anom}.
Eliminating $\rho_{\cal A}$ in terms of $p_{\cal A}$, we are left
with a first order differential equation for $p_{\cal A}$. This
leads to one integration constant which as before contributes to
$T_{\mu\nu}^{\cal A}$ as a `radiation' term of the same form as
\eqref{statedep.T}. For the same reasons of
Sec.~\ref{sec:absence}, this has to vanish and $T_{\mu\nu}^{\cal
A}$ is completely fixed.
Therefore, in the semiclassical approach the quantum corrected
version of the `Friedman' equation \eqref{GR.a} becomes
(concentrating on the $\kappa=1$ case)
\begin{equation}
\label{friedman}%
3\mpl^2\; \frac{1-R'^{\,2}}{R^{\,2}}=\frac{3\,N^2}{32\pi^2}
    \left(\frac{1-R'^{\,2}}{R^{\,2}}\right)^2-P\,,
\end{equation}
where $P$ is the pressure of the DW in the transverse direction.

Rather than finding the exact solutions of \eqref{friedman} given
a smooth profile for $P$, we shall turn to the thin wall
description and restrict our attention to the extrinsic curvature
$K_0$ and the intrinsic curvature $H_0$ on the DW.
Technically, this can be done by integrating across the domain
wall the component of \eqref{semiclassical} that contains the
$\delta$-functions associated with the DW energy density. It can
be easily reproduced from \eqref{friedman} by taking one
derivative with respect to $\xi$ (and dividing by $R'/R$) and
using \eqref{cons}.
Given that the warp factor $R(\xi)$ is continuous, it can be
treated as a constant across the DW core. Using also that
$R'(\xi)$ has a finite discontinuity, and so the integral of any
power of $R'$ does not contribute, we find
\begin{equation}
\label{loop.corr} K_0=\frac{1}{4\mpl^2}\left(\sigma+
\frac{N^2}{12\pi^2}\lp( 3 K_0 H_0^2 - K_0^3\rp) \right)\,,
\end{equation}
Hence, the only effect of the CFT is to renormalize the DW tension
by an amount given in terms of the geometric invariants of the DW
worldvolume, and \eqref{loop.corr} is all we need to solve in
order to find the self-consistent solutions.\\

Let us emphasize that even though this is a 1-loop computation,
this is \emph{all} from the CFT. In general, one expects higher
loop planar diagrams to contain contributions of the same order in
the $1/N$ expansion and with a non-trivial dependence on the 't
Hooft coupling $\lambda=g^2_{YM}N$. Recall that the AdS/CFT
correspondence relates this to the gravity side as
$\lambda=(\ell/\ell_s)^4$ where $\ell_s$ is the string scale and
$\ell$ is the AdS curvature scale. Hence, the classical gravity
description of Sec.~\ref{sec:RS} corresponds to strong coupling
($\lambda\gg1$), so higher loops may not be negligible. However,
in our case the higher loops do not contribute. The reason is
that, as shown above, the CFT effects arise from the trace anomaly
only. In ${\cal N} = 4$ SYM, the trace and the chiral anomalies
are in the same supermultiplet. Because of the non-renormalization
theorem for the chiral anomaly, the trace anomaly is protected as
well. Hence, \eqref{loop.corr} should also be valid for large
$\lambda$,
as the results of Section \ref{sec:RS} will indeed confirm.\\

Equation \eqref{loop.corr} is in its most generic form, and one
could use the Gauss equation \eqref{GaussCodacci} to express it in
terms of $K_0$ and the brane curvature $H_4$. We shall now
restrict to flat brane case, for which $H_0=|K_0|$  and
\begin{equation}\label{cftResult}
\frac{\sigma}{4\mpl^2}=K_0\left[1-\frac{1}{3}(K_0\ell)^2\right]\,.
\end{equation}
Here we introduced
\begin{equation}\label{ell}
\ell^2=N^2/(8\pi^2\mpl^2)~
\end{equation}
merely as a short-hand notation, but as will become apparent in
Section \ref{sec:RS}, $\ell$ is to be identified in the
Randall-Sundrum setup as the curvature scale of the AdS bulk.
Also, as we shall see shortly, $1/\ell$ plays the role of the
cutoff of the theory \cite{ahpr}. This can also be understood
along the lines of \cite{gia07}, as a consequence of the
consistency of the theory and the fact that the CFT contains of
order $N^2$ degrees of freedom coupled to gravity, which forces
the cutoff to be $\mpl/N$ rather than the naively expected $\mpl$.

\begin{figure}[!tb]
\begin{center}
  \includegraphics[height=5cm]{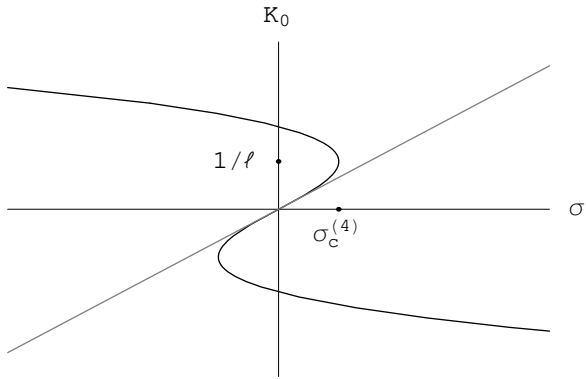}%
  \caption{The extrinsic curvature of the Domain Wall $K_0$ as a
  function of its tension $\sigma$. For small enough $\sigma$,
  there is one branch very
  close to the GR result (light gray). For the other branches,
  $K_0$ is of the order or larger than the cutoff $1/\ell\sim \mpl /N$, and
    they cannot be trusted.
  }
\label{fig:cft} %
\end{center}
\end{figure}

Equation \eqref{cftResult} is cubic in $K_0$, and so it admits up
to three branches, of which the `normal' branch is the one that
reduces to the GR result when the CFT is removed. As illustrated
in Fig \ref{fig:cft}, for tensions below a critical value given by
\begin{equation}\label{sigmac4}
  \sigma^{(4)}_c = {8\over 3} {\mpl^2\over\ell} =
{16\sqrt2\,\pi\over 3} {\mpl^3\over N}
\end{equation}
two new branches of solutions appear. One of them behaves with
negative effective tension (\emph{i.e.}, $K_0<0$) and is thus
expected to be unstable. The other has positive effective tension,
so it may be stable. One intriguing feature of these solutions is
that they display a kind of self-acceleration: even for vanishing
$\sigma$, the quantum corrected DW is inflating, at a rate
entirely due to the CFT radiative corrections. However, just
because the loop corrections are more important than the tree
level effect, this suggests that this solution cannot be trusted.
Indeed, given that the CFT correction is suppressed by $1/\ell$,
one expects that this scale plays the role of the cutoff of the
theory. But this is precisely the curvature scale of  the two new
branches. Thus, their actual presence is not guaranteed.

Hence, we are left only with the normal branch of solutions, the
one where the radiative corrections are subdominant to the
classical term (for small $\sigma$)\footnote{As we shall see in
Sec.~\ref{sec:RS}, in the 5D dual there is only one branch of
solutions. This gives further indication that the
`self-accelerating' branches of \eqref{cftResult} are not realized
in the full theory.}. Note that these corrections make $K_0$, and
hence the gravitational repulsion generated by the DW,
\emph{larger} than in GR. Also, the difference appears only at
nonlinear level in the tension. This will be exactly reproduced in
the Randall Sundrum setup.

Finally, let us note that these solutions cease to exist for
tensions larger than $\sigma_c^{(4)}$. This happens because the
curvature scale $K_0$ becomes of order of the cutoff, $1/\ell$.
Hence, the theory breaks down, that is, the solutions become
sensitive to the UV completion. This is indeed what will see in
the RS setup. For tensions of order $\sigma_c^{(4)}$ or larger,
the walls start to behave in the 5D fashion.

%
%

\section{Domain Walls in Randall Sundrum}
\label{sec:RS}

As argued above, the Domain Walls dressed with the CFT radiative
corrections should be dual to DWs localized on a Randall Sundrum
brane \cite{rs2}. This amounts to finding the metric produced by a
codimension 2 brane (the Domain Wall) embedded in a codimension 1
brane (from now on, \emph{the} brane). This was first discussed in
\cite{gp1}, and here we will closely follow their derivation.

Since one can view the brane as moving with a uniform acceleration
in the bulk, the solution we are looking after will contain an
accelerated codimension 2 brane pulled by the codimension 1 brane.
Hence, one can already expect the solution to be appropriately
described in some kind of Rindler coordinates which, as we shall
see, will appear naturally. As a consequence, and in contrast with
isolated codimension 2 branes, the worldvolume of the accelerated
ones inflates at a rate that depends on their tension. This
already represents a `zeroth order' check of the correspondence,
since Domain Walls in 4D (with or without the CFT)
do inflate at a rate sensitive to $\sigma$.\\


In the Randall Sundrum model, the bulk has a negative cosmological
constant $\Lambda_5$, the brane is characterized by a tension
$\tau$. Including the DW, the action is
\begin{equation}
    S=\int d^5x \sqrt{-g} \;\lp({M^3 \, R_5\over2} -\Lambda_5 \rp)%
    -\int d^4x \sqrt{-h} \;\tau %
    -\int d^3x \sqrt{-\gamma} \;\sigma\,, %
\end{equation}
where $R_5$ is the bulk Ricci scalar, and $h_{\mu\nu}$ and
$\gamma_{\mu\nu}$ are the induced metrics on the brane and on the
DW respectively.

For the time being, we will not assume any relation between the
AdS curvature $\Lambda_5$ and the tension $\tau$ of the brane.
%
The equations of motion for the brane (the Israel junction
conditions) are
\begin{equation}
    \label{israel} %
    2 M^3 k_{\mu\nu}=T_{\mu\nu}-{1\over3}T h_{\mu\nu}\,,
\end{equation}
where we imposed $\Z_2$ symmetry across the brane, $k_{\mu\nu}$ is
the brane extrinsic curvature and $T\equiv T^{\mu}_{\mu}$.
$T_{\mu\nu}$ is the energy-momentum tensor associated with the
brane and the DW,
\begin{equation}
T_{\mu\nu}=-\tau h_{\mu\nu} -\sigma \delta(\xi) \gamma_{\mu\nu}\,.
\end{equation}
As before,  $\xi$ is the (proper) coordinate along the brane which
is orthogonal to the DW. In terms of $\xi$, the position of the DW
is at $\xi=0$.

\subsection{Solution}
\label{sec:sol}

As before, we will concentrate on the solutions with the symmetries
of a maximally symmetric DW, that is with a 3D maximally symmetric
slicing. The (double-Wick rotated version of the) Birkhoff theorem
guarantees that the most generic solution with this symmetry can
be written locally as
\begin{equation}
\label{bulkmetric}
ds^2=f(R)dZ^2 + {dR^2\over f(R)} +R^{2}ds_{\kappa}^2\,,
\end{equation}
where $ds_{\kappa}^2$ is the line element of a 3D maximally
symmetric space of unit curvature radius, that is, a de Sitter
($\kappa=1$), Anti de Sitter ($\kappa=-1$) or Minkowski
($\kappa=0$) spacetime.

The form of $f(R)$ depends on the presence of bulk fields, and in
their absence it is equal to
\begin{equation}
\label{f} f(R)=\kappa+{R^2\over\ell^2}+{\mu\over R^2}\,.
\end{equation}
Here, $\ell^2=|6M^3/\Lambda_5|$  is the AdS curvature radius, and
$\mu$ an integration constant. This solution is a double Wick
rotation of the Schwarzschild-AdS metric, \emph{i.e.}, the AdS
bubble of nothing
\cite{WittenBON}--\nocite{afhs,birmingham,balaRoss}\cite{balaKlausSimon}.

In terms of these `bulk adapted' coordinates, the full spacetime
can be constructed as usual by finding the embedding of the brane
in the bulk (which is determined by the junction conditions
\eqref{israel}), cutting across the brane location and gluing two
copies of the bulk along the brane.

One can always parameterize the location of the brane by two
functions \linebreak $(R(\xi),Z(\xi))$, and solve for them by
imposing that the Israel junction conditions are satisfied. A
level of arbitrariness is still present, due to the
re-parametrization (gauge) invariance of the embedding. To fix the
gauge, it is convenient to choose
\begin{equation}\label{Z}
f(R)Z'^2+{R'^2\over f(R)}=1\,.
\end{equation}
With this condition, the induced metric on the brane precisely
takes the form \eqref{induced}, and $\xi$ is the proper distance
on the brane perpendicular to the DW. Equation \eqref{Z} relates
$Z(\xi)$ in terms of $R(\xi)$, so once $R(\xi)$ is known, the
embedding of the brane in the AdS bulk is determined.

In the following, we shall find the form of $R(\xi)$ in the thin
wall approximation and place the wall at $\xi=0$. With this in
mind, we can solve the components of \eqref{israel} along the
three-dimensional slicing firstly away from the wall,
\begin{equation}
\label{angular}%
2\,{\rm sign}(\tau) \;M^3\; {\sqrt{f(R)-R'^2}\over  R} =
{\tau\over3}\,. 
\end{equation}
One can always split the tension as $\tau=\tau_{RS}+\delta\tau$,
with
\begin{equation}
    \tau_\mathrm{RS}\equiv\frac{6M^3}{\ell}\,.
    \label{tauRS}
\end{equation}
Then, \eqref{angular} leads to the analogue of the Friedman
equation
\begin{equation}
\label{angular2}%
3\mpl^2\frac{\kappa-R'^2}{R^2}= 
\Lambda_4^\eff - {3\mu  \mpl^2\over R^4}%
 \,, 
\end{equation}
where we used that $M^3\ell=\mpl^2$ and we identified the
effective 4D cosmological constant as
\begin{equation}\label{lambda4eff}
  \Lambda_4^\eff=\delta\tau\lp(1 + {\delta\tau\over
  2\tau_{RS}}\rp)~.
\end{equation}
Choosing the brane tension equal to the critical value $\tau_{RS}$
is the the
so-called `Randall-Sundrum condition', and gives $\Lambda_4^\eff=0$.\\

%

%

We are now ready to see that the DW motion generates no particle
production in this setup. The `dark radiation' term (the last one
on the right hand side of \eqref{angular2}) maps to the state
dependent contribution \eqref{statedep.T} of Sec.~\ref{sec:cft},
which encodes the particle creation effects. Accordingly,
$3\mu\mpl^2$ is mapped to $p_0$. In Sec.~\ref{sec:cft}, we
concluded that $p_0=0$ as long as the DW is maximally symmetric
and there is a horizon. Let us now see that in the same
circumstances, one also concludes that $\mu=0$ in the RS setup.

As before, we will restrict our attention to $\kappa\neq-1$, which
is when there is a horizon. For $\mu>0$ the bulk has a naked
singularity at $R=0$, so this case is unphysical. For $\mu<0$,
instead, $R$ becomes a radial coordinate with center at the zero
of $f(R_0)=0$, and the bulk is a smooth space as long as $Z$ is
periodically identified, $Z\simeq Z +4\pi /f'(R_0)$. When solving
for the brane trajectory from \eqref{angular2}, one can see that
$R(\xi)$ bounces and does not vanish anywhere, \emph{i.e.}, there
is no horizon. One can also show by integrating \eqref{Z} that
$Z(\xi)$ grows unbounded (for any value of $\Lambda_4$). But given
that $Z$ is compact, this implies that $\xi$ must be compact as
well. As we noted in Sec.~\ref{sec:cft}, solutions with $p_0<0$
could be expected for a compact $\xi$, but not otherwise since
then the curvature scale is of order of the cutoff at the bounce.
In the 5D dual, we see that indeed only the solutions with a
compact $\xi$ are present.

Hence, the only case with a regular bulk and a Rindler horizon on
the brane is when $\mu=0$. This argument, based entirely on the
geometry, is the 5D dual of the argumentation that led us to
conclude in the CFT side that there is no particle production.
Hence, from now on we will set $\mu=0$. We will comment further on
the geometrical properties of the space \eqref{bulkmetric} with
$\mu=0$ in Sec.~\ref{sec:structure}.\\

Now, let us consider the junction  condition at the DW location.
On solving for $R(\xi)$, we have to impose it to be continuous
across the DW, but not its $\xi$-derivatives. The reason for this
can be easily seen from the $(\xi,\xi)$ component of
\eqref{israel}, which near the localized DW reads
\begin{equation}
\label{xixi1}%
-2\, {\rm sign}(\tau)  \,M^3 {R''\over \sqrt{f(R)-{R'}^2}} =\sigma
\;\delta(\xi)\,.
\end{equation}
By integrating across the DW, we obtain the matching condition for
the discontinuity on $R'$, namely\footnote{Note that this was
incorrectly derived in \cite{gp1}.}
\begin{equation}
\label{xixi}
-\,{\rm sign}(\tau)\,\Delta\arctan{R'_0\over\sqrt{f(R_0)-{R'_0}^2}}=\frac{\sigma}{2M^3}
\end{equation}
where $\Delta X\equiv(X(0^+)-X(0^-))$, and it should be noted
that, away from the DW the $(\xi,\xi)$ component of \eqref{israel}
is automatically satisfied by \eqref{angular}.

\begin{figure}[!tb]
\begin{center}
  \includegraphics[height=5cm]{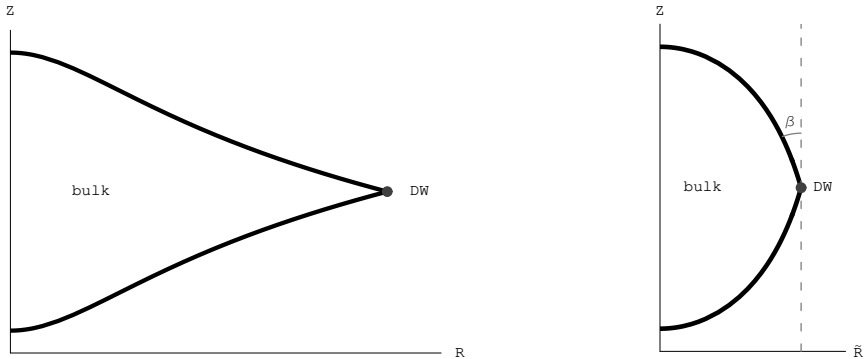}%
\end{center}
\caption{A Domain Wall (the thick dot) localized on the brane (the
thick line). The full space consists of two copies of the bulk
glued along the brane. The left panel is in the $R$ coordinate,
which represents the Hubble radius of the 3D sections (here
suppressed). The right panel is in the conformal coordinates, $Z$
and $\tilde R=\ell\arctan(R/\ell)$, in terms of which we can
easily identify the deficit angle $\delta$. For positive brane
tension $\tau$, the bulk corresponds to the interior, so
$\delta=4\beta$. For $\tau<0$, the bulk is the exterior and
$\delta =-4\beta$. 
}
\label{fig:2dw-diwall-dirs2} %
\end{figure}

This condition is better understood introducing the ``conformal''
coordinate $\tilde R=\ell\arctan(R/\ell)$, in terms of which the
$ds^2_{\kappa}=0$ sections of the metric take the form
$ds^2=f(\tilde R)(dZ^2+d\tilde R^2 )$. In these coordinates, the
angle of the brane trajectory (with respect to the $Z$ axis) is
\begin{equation}
\label{beta}%
\tan\beta\equiv{\tilde R'\over Z'}= {R'\over \sqrt{f(R)-R'^2}}~.
\end{equation}
One can see from Fig. \ref{fig:2dw-diwall-dirs2} that the deficit
angle in the solution is given by
$$
\delta =2\,{\rm sign}(\tau)\,\Delta\beta=4\,{\rm sign}(\tau)\,
\beta|_{0^+}
$$
Hence, Eq.~(\ref{xixi}) is the statement that the DW generates a
deficit angle given by
\begin{equation}
\label{deficit1} %
\delta={\sigma\over M^3}~.
\end{equation} %
This is the usual relation between the tension and the deficit
angle of a codimension 2 object.
In this sense the DW gravitates completely as expected from the 5D
point of view.\\

Now, let us see how this looks from the point of view of an
observer living on the brane. Using (\ref{angular}), one easily
obtains that the junction condition on the wall (\ref{xixi})
translates into
\begin{equation}\label{junctionHere} %
K_0 = {\tau \over 6 M^3} \tan\lp( {\sigma \over 4 M^3}\rp)~,
\end{equation} %
where $K_0$ is as in \eqref{K0}. We see that the jump in the DW
extrinsic curvature ({\em i.e.}, its accelerated motion) is
determined by the tensions of both the codimension 1 and the
codimension 2 branes. This is expected since for $\tau=0$ the DW
should not be accelerating, as any genuine codimension 2 brane.

Note as well that \eqref{junctionHere} together with
\eqref{angular2} (with $\mu=0$) and \eqref{GaussCodacci} lead to
the following relation for the Hubble rate with respect to the
brane tensions and bulk cosmological constant,
\begin{equation}\label{H0here}
H_0^2 = \lp({\tau \over 6 M^3}\rp)^2 {1\over  \cos^2\lp( {\sigma
/4 M^3}\rp)} + {\Lambda_5\over 6 M^3}~.
\end{equation}
This can be viewed as the effective `Friedmann equation' valid on
the DW. As expected, for $\tau=0$ (an isolated codimension 2
brane), this trivially sets the Hubble rate equal to that of the
bulk. However, in the presence of the codimension 1 brane, $H_0$
becomes sensitive to the energy density $\sigma$.\\

In summary, we have seen that the induced metric on the brane
takes exactly the same form as in GR (a VIS spacetime or its
generalizations with nonzero cosmological constant) except for the
relation between $K_0$ (or $H_0$) and the tension $\sigma$. We
leave for Sec.~\ref{sec:comparison} the comparison between this
result and the one obtained in the CFT
description \eqref{cftResult}.\\

At this point, the first comment is that the linearized version of
the exact result \eqref{junctionHere} correctly reproduces what
one would find in the linear theory \cite{gt}. Indeed for $\tau >
0$, the KK decomposition of the 5D graviton includes a zero mode
coupled with an effective 4D Plank mass given by $\ell M^3$
together with a tower of massive gravitons. The latter do not
couple to relativistic DWs \cite{dgpr}, so the only effect comes
from the zero mode, precisely matching \eqref{junctionHere} at
linear level. For $\tau<0$ (and a single brane), the spectrum
consists of a scalar zero mode (the radion), the massive KK
gravitons and no graviton zero mode. Hence, the effect comes from
the radion, which `explains' the opposite sign in $K_0$.

Another important comment is that both $K_0$ and $H_0$ diverge
when $\sigma$ approaches the critical value
$$
\sigma_c^{(5)}=2\pi M^3~,
$$
that is, when the deficit angle saturates to $2\pi$. This is an
unavoidable feature of accelerated codimension 2 branes and the
reason can be traced back to the Gauss-Bonnet theorem applied to
the (Euclideanized) space transverse to the DW. The theorem states
that the Euler characteristic $\chi$ is given by
\begin{equation}\label{GB}
2\pi\chi=\int_{\cal M} \sqrt{g_2} d^2x R_2 + \int_{\partial\cal M}
\sqrt{g_1} dx K
\end{equation}
where $R_2$ is the Ricci scalar of the space $\cal M$ transverse
to the DW (with metric $g_{2}$) and $K$ is the extrinsic curvature
at the boundary $\partial \cal M$. Since $\chi$ is a topological
invariant, in particular it must be independent of the DW tension.
Using the equations of motion, one finds a contribution from the
DW tension (from the boundary integral). Away from the DW, the
local curvatures $R_2$ and $K$ are constant and independent of
$\sigma$. So, the only way that the integral is independent of
$\sigma$ is that the integration volume changes. In our case, the
topology of the transverse space is that of a disk, so $\chi=1$.
When the DW tension approaches $2\pi M^3$, its contribution to the
right hand side of \eqref{GB} saturates to $2\pi$. Hence, the
remaining `volume' contribution has to vanish, which implies that
$H_0^{-1}$ must vanish too.

This is of course related to the `pathology' present for isolated
codimension 2 branes. The opening angle of the conical transverse
space is zero for critical tension, and the two transverse
directions `collapse' to a line. If one resolves the string
thickness, however, one realizes that the transverse space is a
cylinder with a radius of order of the thickness. Similarly, the
divergence in $H_0$ and $K_0$ in the present case will be resolved
by introducing the wall thickness. We defer this discussion to
Section \ref{sec:RSthickness}.

\subsection{Beyond the horizon}
\label{sec:structure}

Let us briefly comment on the properties of the bulk spacetime and
especially on continuation through the horizon, with the purpose
of showing explicitly that no pathologies, like Closed Timelike
Curves (CTCs), are present. Specifically, we shall concentrate on
the case $\kappa=1$.

\begin{figure}[!tb]
\begin{center}
  \includegraphics[width=6cm]{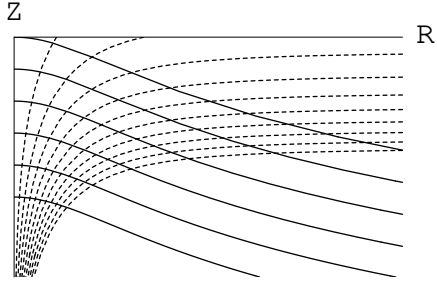}%
\caption{Contours of constant $y$ coordinate (solid line) and
constant $r$ coordinate (dashed) in the $R,\,Z$ plane. $y$ is the
proper coordinate in the Poincar\'e slicing of AdS${}_5$ (see
\eqref{poincare}). The dashed lines are (spacelike) geodesics, and
measure the proper distance in the direction perpendicular to the
brane. The solid lines correspond to the location of a `tuned' RS
brane, with a tension given by \eqref{tauRS}. 
}
\label{fig:RZ_ry} %
\end{center}
\end{figure}

The bulk of the above solution is locally AdS${}_5$, which in the
$R,\,Z$ coordinates takes the form
\begin{equation}\label{ads}
    ds^2=\left(\kappa+\frac{R^2}{\ell^2}\right)dZ^2+
    \left(\kappa+\frac{R^2}{\ell^2}\right)^{-1}dR^2+R^2\;ds_\kappa^2\,.
\end{equation}
These coordinates are convenient because the DW sits at a point in
the $R,\,Z$ plane, and its Hubble radius is simply given by the
value of the $R$ coordinate. While we will keep part of our
discussion for generic $\kappa$, the case of most interest for us
is going to be $\kappa=1$, when the DW inflates. In this case, the
above coordinates do not cover all of AdS${}_5$ because $R$ is a
kind of Rindler coordinate. This can be readily seen from the form
of the coordinate transformation that brings $R,\,Z$ to the
Poincar\'e patch,
\begin{eqnarray}\label{poincare}
R&=&r \, e^{-y/\ell} \cr%
e^{2Z/\ell}&=&e^{2y/\ell}+(r/\ell)^2.
\end{eqnarray}
In terms of these, the metric is $ds^2=dy^2+ e^{-2 y\ell} ( dr^2 +
r^2 ds_{(\kappa=1)}^2)$, so we identify $r$ as the usual Rindler
coordinate of the 4D Minkowski sections (given by constant $y$
slices). Figure \ref{fig:RZ_ry} depicts the relation between these
coordinates. The constant $r$ lines are geodesics, while the
constant $y$ are curves with uniform acceleration which precisely
correspond to the location of a tuned RS brane (see below).\\

Now, let us discuss the continuation across the horizon. Starting
from the metric \eqref{ads} and writing the line element of the 3D
sections as $-d\tau^2 + \cosh^2\tau\,d\Omega^2$, where $d\Omega^2$
is the metric on the 2-sphere,  the continuation across the
Rindler horizon at $R=0$ is given by $R\to i T$ and
$\tau\to\chi+i\pi/2$. Thus, the metric in the Milne region looks
like
\begin{equation}\label{milne1}
    ds^2=\left(1-\frac{T^2}{\ell^2}\right)dZ^2-
    \left(1-\frac{T^2}{\ell^2}\right)^{-1}dT^2+T^2\;dH_3^2\,,
\end{equation}
where $dH_3^2=d\chi^2+\sinh^2\chi d\Omega^2$ is the metric on the
hyperbolic space. These coordinates are again singular at
$T=\ell$. Hence, one has to do one more continuation, $Z\to
\widetilde T + i\pi/2$, $T\to X$, and in this patch (with
$X>\ell$) the metric takes the form
\begin{equation}
    ds^2=-\left(\frac{X^2}{\ell^2}-1\right)d\widetilde T^2+
    \left(\frac{X^2}{\ell^2}-1\right)^{-1}dX^2+X^2\;dH_3^2\,.
\end{equation}
Thus, $Z$ becomes a time coordinate when the radius of the 3D
slices becomes larger than $\ell$.

In these coordinates, it is apparent how the BTZ black hole
\cite{btz,bhtz} (and its higher dimensional generalization)
arises. This is obtained simply by periodically identifying $Z$,
which implies a periodic $\widetilde T$. In order not to have
CTCs, one has to excise the region covered by $X\,,\widetilde T$.
Then, the hypersurface $T=\ell$ plays the role of a singularity in
the sense that geodesics terminate there.

Even though we did not make it manifest until now, we are assuming
that $Z$ has a noncompact range (for $\kappa\neq-1$). At first,
this seems to guarantee that the bulk is pure AdS${}_5$ as opposed
to a BTZ black hole. However, since we are doing a nontrivial
identification by gluing two copies of the bulk by the brane
location, we should make more explicit that this does not give
rise to CTCs. In particular, this means that the brane trajectory
must be such that its $Z$ coordinate goes to infinity when
approaching $T=\ell$, otherwise the bulk contains CTCs with finite
period. This turns out to be precisely what happens.

\begin{figure}[!tb]
\begin{center}
  \includegraphics[height=5cm]{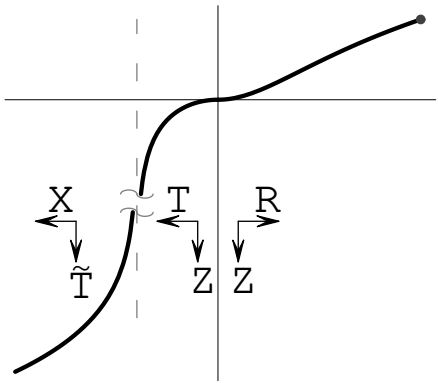}%
\end{center}
\caption{Form of the brane trajectory in the bulk. Only one `arm'
(one side of the DW) of the brane is shown. The continuous
vertical line corresponds to the Rindler horizon. To the right of
it, there is the Rindler region; to the left, there is the the
Milne region,covered by the patches $T\,,Z$ and $X\,,\widetilde
T$. The dashed axis corresponds to $T=X=\ell$, and is where $Z(t)$
and $\widetilde T(t)$ diverge.}
\label{fig:continuation} %
\end{figure}

For the tuned RS case, the induced metric on the brane takes the
form of the VIS model described in Sec.~\ref{sec:GR}. In the
Rindler region, this is given by $R(\xi)=H_0^{-1}-|\xi|$. The
continuation across the horizon gives a Milne universe, $-dt^2+t^2
dH_3^2$ and the conformal diagram on the brane is the same as Fig
\ref{fig:VIS}. Hence, we can identify the trajectory in the $T\,,
Z$ plane as $T(t)=t$ and $Z(t)$ as given by the continuation of
\eqref{Z}. Similarly, in the region covered by the $X\,,\widetilde
T$ coordinates, one has $X(t)=t$. One can easily see that
$Z(t)=(1/2)\log(1-(t/\ell)^2)$, so indeed $Z$ (and similarly
$\widetilde T$) diverge for $T$ close to $\ell$. This is
illustrated in Figure \ref{fig:continuation}, where we show the
form of the brane trajectory in a combined $(R\,,Z)$, $(T\,,Z)$
and $(X\,,\widetilde T)$ plane.

In summary, there are no CTCs in the region covered by $X\,
,\widetilde T$ so there is no need to excise this region.
Ultimately, this means that the continuation of the brane geometry
includes the whole Milne region and the conformal diagram is
indeed given by Fig. \ref{fig:VIS}.

\subsection{Beyond the critical tension}
\label{sec:RSthickness}

In this Subsection, we shall see how the divergence of $H_0$ and
$K_0$ when the DW tension reaches the critical value $2\pi M^3$ is
resolved by introducing the DW thickness, $d$. As a byproduct,
this allows us to find solutions for supermassive codimension 2
branes. In the next derivation, we will follow \cite{dgpr}.

As it happens also in GR (see Section \ref{sec:GR}), once we
introduce the DW thickness, one has to allow for a nonzero
pressure in the direction orthogonal to the wall, $P$. Hence, the
form of the stress tensor for the DW is as in \eqref{tmunu}. We do
not need to specify the actual microscopic model for the domain
wall, but we shall assume that its stress tensor is conserved on
the brane, which leads to \eqref{cons}.

From \eqref{resultRS} and \eqref{H0here}, as soon as the DW
tension $\sigma$ approaches $2\pi M^3$ both $H_0$ and $K_0$ grow
very quickly beyond $1/\ell$. From the Gauss Codazzi equation
\eqref{GaussCodacci}, one also has that $H_0\sim |K_0|$ (for a
moderate brane tension). In addition, we shall assume that the DW
thickness $d$ is much smaller than $\ell$. In this case, there is
a range of tensions for which
$$
1/\ell \ll H_0 \ll 1/d~.
$$
In this regime, the DW is not yet in a phase of topological
inflation\footnote{As in Sec.~\ref{sec:GR}, by topological
inflation we mean that the Hubble radius on the wall is smaller
than the thickness. If so, in the core of the DW inflation takes
place in all directions.}, and we can still unambiguously speak of
the DW. The interest of this range is that the DW experiences 5D
gravity.

We can proceed essentially along the same steps as in the previous
Subsection, but now including the pressure $P$
\begin{equation}
\label{with pressure}%
2\,M^3\; {\sqrt{f(R)-R'^2}\over  R} = {\tau-P\over3}~,
\end{equation}
For simplicity we will focus on the case $\tau=\tau_{RS}>0$ and we
shall also assume that $P$ is everywhere less than $\tau$, so that
the sign of the right hand side is positive.

Instead of computing explicitly the other component of the
junction conditions, we can derive it as in Section \ref{sec:cft},
from \eqref{with pressure}.
One obtains
\begin{equation}
\label{xixiThickAllTerms}%
-2\, M^3 \lp\{ {R''\over \sqrt{f(R)-{R'}^2}} + {1-R'^2\over R
\sqrt{f(R)-{R'}^2}} \rp\}=\rho+ P.
\end{equation}
As before, the idea is to integrate across the DW core this
expression and obtain a junction condition on the DW. This time,
we shall not only keep the $\delta$-function-like terms, since we
want to keep track of the contribution from $P$. We must also take
into account that the second term on the left hand side of
\eqref{xixiThickAllTerms} is of the same order as $P$. Using
\eqref{with pressure}, we arrive at
\begin{equation}
\label{xixiThick}%
-2\, M^3 {R''\over \sqrt{f(R)-{R'}^2}} =\rho+{1\over3} P.
\end{equation}
where we have ignored terms of order $P^2/\tau\ll P$.

\begin{figure}[!tb]
\begin{center}
  \includegraphics[height=5cm]{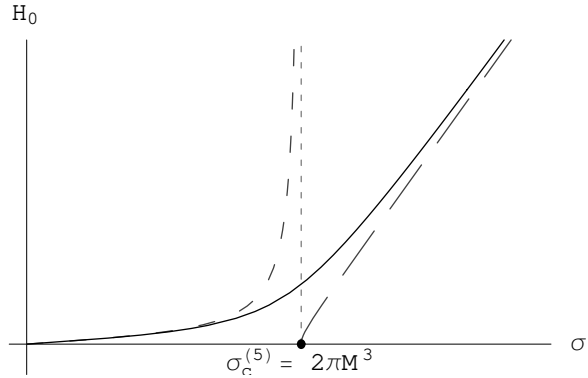}%
\caption{Relation between the DW Hubble rate $H_0$ and its tension
$\sigma$ as given by \eqref{thickRS} for $\ell/d=10$ (solid line).
The singularity at $\sigma=\sigma_c^{(5)}$ in the thin wall limit
(short dashed line) gets resolved. The long dashed line
corresponds to a deficit angle equal to $2\pi$. As is apparent,
even for supercritical tensions ($\sigma>\sigma_c^{(5)}$), the
deficit angle is never saturated. } \label{fig:thickRS}
\end{center}
\end{figure}

Integrating across the DW, one obtains the following matching
condition
\begin{equation}
\label{matchThick}%
 \frac{1}{2M^3}\lp(\sigma+
 {1\over 3}\,d\,P_0\rp)%
= -\, \Delta\;\arctan{R'\over\sqrt{f(R)-{R'}^2}}  %
\simeq 2 \, \arctan{\ell\over R_0}
\end{equation}
where in the second equality we used that $H_4$ is much smaller
than $K_0,H_0$ and that in the core of the DW $R(\xi)$ stays close
to its central value, $R(0)=1/H_0$. Also, we identified the DW
tension as $\sigma= \int d\xi \rho$,  and we defined the DW
thickness by
$$
d\equiv {1\over P_0} \int_{core} d\xi P~.
$$
From \eqref{with pressure}, the pressure at the center of the core
is given by
\begin{equation}
{P_0\over 6 M^3}= {1- \sqrt{1+(H_0\ell)^{2}}  \over\ell}~.
\end{equation}
Putting these together, we find that $H_0$ now depends on $\sigma$
as
\begin{equation}\label{thickRS}
  \arctan\lp(H_0\ell\rp)+ {1\over2}{d\over \ell } \lp(
\sqrt{1+(H_0\ell)^2} -1\rp) ={\sigma \over 4 M^3}~.
\end{equation}
We plot this relation in Fig. \ref{fig:thickRS}, where it is
apparent that the singularity at $2\pi M^3$ in \eqref{H0here} and
\eqref{junctionHere} is resolved. Note that even though we can
have $\sigma > 2\pi M^3$, the deficit angle (which is given by the
first term in the left hand side of \eqref{thickRS}) never exceeds
$2\pi$.

As expected, the contribution from $P$ in \eqref{thickRS} vanishes
for $d\to0$. However, this term allows us to have supercritical
solutions. Indeed, for $\sigma\gg 2\pi M^3$ one has $H_0 \ell \gg
1$ and \eqref{thickRS} approximates to
\begin{equation}\label{superduper}
  H_0 \simeq {\sigma - 2\pi M^3 \over 2 d M^3}~.
\end{equation}
This is similar to a usual DW with an effective 4D Planck mass of
order $d M^3$. So, in this regime, the gravitational effect of the
DW is enhanced by a factor $\ell/d$. The change of behaviour
always takes place for $\sigma$ close to $2\pi M^3$ (and
consequently $H_0$ close to $1/\ell$).

This new regime should be identified as the five dimensional
behaviour not just because $H_0 \gg 1/\ell$, but also because this
is how a (supermassive) codimension 2 branes behave \cite{dgpr}.
Indeed, for subcritical tensions, codimension 2 branes do not
inflate, rather they only produce a deficit angle given by
$\sigma/M^3$. For critical tension, the transverse space becomes a
cylinder of radius of order $d$. Hence, the space is effectively
compactified to one dimension less\footnote{In our case, the
compactification is effective only locally. A distance of order
$\ell$ away from the wall, the brane trajectory is similar to the
subcritical case, and the bulk opens up.}, with an effective
Planck mass of order $d M^3$. For supercritical tension, there are
no regular static solutions \cite{gott,ortiz,lg}, but inflating
ones may exist \cite{deLaix,cho}. This is perfectly compatible
with the fact that the transverse space has been compactified, so
the brane effectively behaves as a codimension 1 object from the
lower dimensional point of view. Moreover, the brane `spends' part
of its tension ($2\pi M^3$ to be precise) compactifying, and the
remainder in how much it inflates. Hence, the gravitational effect
of a supermassive codimension 2 is to inflate precisely according
to \eqref{superduper}, and we conclude that the DW is exhibiting a
5D behaviour as opposed to a 4D one (which would give $H_0\simeq
\sigma/4\mpl^2$).

We should add that once the DW experiences this 5D gravity,
$\sigma$ does not need to be much larger than $2\pi M^3$ before
topological inflation sets in. Indeed, for $\sigma\simeq 2(\pi+1)
M^3$ one already has $H_0 \simeq 1/d$ in \eqref{superduper}.

This behaviour has simple interpretation in terms of the
microscopic model for the topological defect. We have in mind a
scalar field model with a quartic potential, but this discussion
should be rather generic. As it turns out, both for Domain Walls
\cite{vilenkinTI} an cosmic strings \cite{deLaix,cho}, the
topological inflation starts when the vev of the scalar field
$\la\phi\ra$ (the location of the degenerate minimum) is of order
of the Planck mass. What happens with the localized DWs is that
they can be viewed both as a codimension 1 and a codimension 2
objects, and the associated Planck scales for each behaviour are
different. As a result, from the 4D point of view, topological
inflation is `prematurely' reached  for $\la\phi\ra$ of order $M$,
rather than the naively expected $\mpl$.

\subsection{Comparison with the CFT}
\label{sec:comparison}

Let us now see how the results of the previous subsection are in
agreement with the dressed DW picture of Section \ref{sec:cft}.
Starting by the most general case, when the brane tension is not
tuned to $\tau_{RS}$, we set $\tau=\tau_{RS}+\delta\tau$ with
$\tau_{RS}$ defined in (\ref{tauRS}). We can rewrite
(\ref{junctionHere}) as
\begin{equation}\label{resultRS} %
{1\over \ell} \arctan\lp( {K_0\ell\over
1+\delta\tau/\tau_{RS}}\rp) = {\sigma \over 4 \mpl^2}~.
\end{equation} %
where we used $\mpl^2=\ell M^3$. Expanding to the leading order
correction, one obtains
\begin{equation}\label{expanded} %
K_0 \lp( 1 - {\delta\tau\over\tau_{RS}} - {1\over3}(K_0\ell)^2
+\dots\rp) = {\sigma \over 4 \mpl^2}~. 
\end{equation} %
From \eqref{angular2} (with $\mu=0$), the curvature scale on the
brane away from the DW is $H_4^2 =\Lambda_4^\eff / 3 \mpl^2$, with
$\Lambda_4^\eff$ given in \eqref{lambda4eff}.  To leading order in
$\delta\tau$, one can rewrite this as $\delta\tau/\tau_{RS} =
(H_4\ell)^2/2 $.
Using this and the Gauss Codazzi equation \eqref{GaussCodacci}, we
find
\begin{equation}\label{powers} %
K_0 \lp( 1 - {1\over2}(H_0\ell)^2 + {1\over6}(K_0\ell)^2
+\dots\rp) = {\sigma \over 4 \mpl^2}~. 
\end{equation} %
This is exactly the same as \eqref{loop.corr} once we use
\eqref{ell}. For the tuned RS case, $\delta\tau=0$, and it is
straightforward that \eqref{expanded} leads to \eqref{cftResult}.\\

Hence, we explicitly see that the `cutoff' AdS/CFT correspondence
works at the level of matching the correct numerical factors. As
mentioned before, this is due to the fact that in the CFT side the
only effects are due to the anomaly, which guarantees that the
results at small 't Hooft coupling (which is what we compute in
the 4D side) and at strong coupling (which is what the gravity
side gives) must coincide. Note that this is not true for the case
that we left out of our analysis, when the DW worldvolume is
AdS${}_3$, because the state dependent contribution from the CFT
\eqref{statedep.T} may be nontrivial.\\

As the reader might have noticed already, the results in both
sides of the correspondence \eqref{cftResult} and \eqref{resultRS}
are \emph{exact}, and yet we find agreement in the first two terms
of the expansion in powers of $K_0,H_0$ (see \eqref{powers}) only.
One can ask what these corrections correspond to in the CFT. These
terms are of the form $K_0 (N^2 K_0^2 /\mpl^2 )^n$ for
$n=1,2\dots$ However, the loop corrections can be arranged as a
$1/N^2$ expansion and the leading order is proportional to $N^2$.
So, these corrections do not come from higher (CFT or graviton)
loops. Rather, these are terms that vanish when the cutoff is
removed, as it is manifest since they are suppressed by the
cutoff, $1/\ell$. Hence from the point of view of the 4D effective
theory they are un-calculable and were implicitly set to zero in
Section \ref{sec:cft}. In the particular UV completion of the 4D
theory provided by the RS setup, instead, they are organized
according to the expansion of \eqref{resultRS}.

Still, the fact that the 4D analysis breaks down for
$\sigma>\sigma_c^{(4)}$ (with $\sigma_c^{(4)}$ given in
\eqref{sigmac4}) signals that the UV completion should change
regime at around that scale. As we have seen in Section
\ref{sec:RSthickness}, this is indeed the case in the RS setup as
the DW starts behaving in a 5D fashion for tensions around
$\sigma_c^{(5)}$, which is parameterically close to
$\sigma_c^{(4)}$.

\section{Conclusions}
\label{sec:conclusions}

\begin{figure}[!tb]
\begin{center}
  \includegraphics[height=5cm]{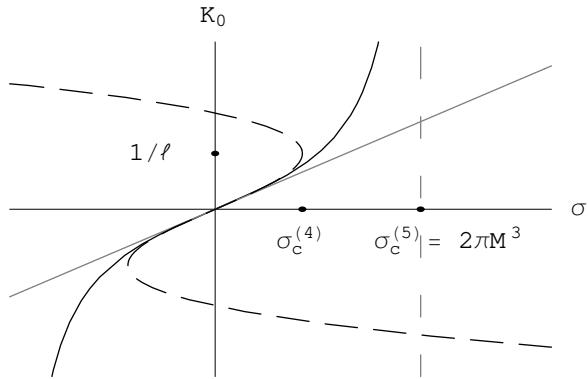}%
\caption{This plot summarizes the comparison between the two sides
of the correspondence. The solid line is the RS result
\eqref{junctionHere}, the dashed line is the outcome from the CFT
side \eqref{cftResult}, for vanishing cosmological constant. We
also include for reference the GR result (solid gray line). For
small enough tensions, the curves from the RS and the CFT results
agree beyond the GR result.
}
\end{center}
\end{figure}

The cutoff version of the AdS/CFT correspondence is a powerful
tool to learn how a (strongly coupled) CFT behaves in the presence
of gravity. In this respect, one of the most relevant problems
that are still being debated is how the black hole evaporation
takes place in this setup and, correspondingly, whether static
solutions for black holes localized on the brane exist
\cite{takahiroBH,efk,frw,FabbriProcopio}.
In this paper we have studied a much simpler case, namely how the
CFT responds to the gravitational field produced by a Domain Wall
(DW). This is a physical implementation of a moving mirror so,
generically, one may expect the presence of particle creation.
Hence, this case represents  a `toy' example where phenomena
similar to Hawking radiation may arise.

We have allowed for an arbitrary cosmological constant
$\Lambda_4$, and we have restricted ourselves to the case when the
DW worldvolume is maximally symmetric.
For $\Lambda_4\geq0$, the DW always develops a horizon while for
$\Lambda_4<0$ it only does so if its tension is large enough.
Whenever the horizon is present, the conformality of the field
theory is enough to see that there is no particle creation.
Given that this follows from very general assumptions concerning
the symmetry of the configuration and the regularity of the
vacuum, this result should hold to all orders in the loop
expansion. The computation in the 5D dual (which is valid at large
't Hooft coupling) confirms that this is the case.
The absence of particle production can also be understood from the
classic results for moving mirrors. When the DW is maximally
symmetric, its motion has a constant acceleration, it is well
knonw that in this case there is no radiation into conformal
fields.

However, the CFT still leads to some effect because the trace
anomaly plays a nontrivial role. This is non-zero on the DW
worldvolume and effectively renormalizes the tension. Hence, there
is a correction to the gravitational field produced by the DW. We
computed it explicitly both in the CFT (at 1-loop) and in the
Randall-Sundrum sides of the correspondence, and find a precise
agreement to the numerical factors. The reason for the numerical
match is that the correction is entirely due to the conformal
anomaly, which has no contributions beyond 1-loop in an ${\cal
N}=4$ SYM.

In the RS side, the computation involves finding solutions for a
DW localized on the brane, that is for a codimension 2 brane (the
DW) embedded on the RS brane. Because of the localization, the
codimension 2 brane behaves very differently from how it would in
isolation. The RS brane `pulls' the codimension 2 brane which, as
a result, moves with uniform acceleration. Hence, it effectively
generates a repulsive gravitational field and, from the point of
view of the observers on the brane, looks like a Domain Wall.

As already emphasized, the amount of symmetries of the case
studied here precludes the DW from radiating CFT quanta. Hence,
there is little that we can add to the black hole debate. However,
our analysis can be extended in several directions ({\em e.g.},
assuming less worldvolume symmetry, a different equation of state
on the wall or including other fields in the bulk), some of which
may lead to particle creation and still be tractable. One
promising case is offered by DWs in AdS${}_4$ when the tension is
low enough so that the worldvolume is AdS${}_3$. In this case, the
wall is still in an accelerated motion but there is no horizon, so
there is no obstruction for particle production. We shall report
on this case soon.

\section*{Acknowledgements}

We thank Gia Dvali, Roberto Emparan, Gregory Gabadadze, Nemanja
Kaloper, Matt Kleban, Michele Redi and especially Massimo Porrati
and Takahiro Tanaka, for useful discussions. This work is
supported by Graduate Students funds provided by New York
University (LG) and by DURSI under grant 2005 BP-A 10131 (OP).

\appendix
\section{1-Loop correction to the graviton propagator}
\label{1loop} In Section \ref{sec:cft}, we described how to obtain
the correction the gravitational field of a DW generated by the
CFT. Using the symmetries of the problem and the properties of the
CFT we concluded that there would be no correction were it not
for the trace anomaly. We shall now show this more directly, by
computing explicitly the first loop correction to the graviton
propagator due to the CFT, and the linearized field that this
gives entails. As we shall see, the 1-loop contribution to the
metric perturbation is pure gauge, so gravitational field is not
affected by the CFT radiative correction. This derivation does not
capture the anomalous term because it is expanded around flat
space, where the anomaly is zero. In this appendix, we shall
follow Ref. \cite{DuffLiu}.

The metric fluctuation around flat space
$h_{\mu\nu}=g_{\mu\nu}-\eta_{\mu\nu}$ can be expressed in momentum
space as
\begin{eqnarray}
\hat{h}_{\mu\nu}&=&-16\pi\GN\,\Delta_{(4)}(p)
    \left[\Tdw_{\mu\nu}-\frac{1}{2}\eta_{\mu\nu}\Tdw\right]+\nonumber\\
    &&-16\pi\,\GN\left[2\Pi_2(p)\Tdw_{\mu\nu}+\Pi_1(p)\eta_{\mu\nu}\Tdw\right]\,,
\end{eqnarray}
where the first term $\Delta_{(4)}$ is the tree level propagator
in four dimensions, while the second is the graviton self-energy
at first loop, and only conformal fields are running in the loop.
The form factors $\Pi_i$ are dictated by symmetries of CFT, and
have the following explicit form
\begin{equation}
\Pi_i=32\pi\GN\,\left[a_i\,\ln\frac{p^2}{\mu^2}+b_i\right]\quad, i=1,2\\
\end{equation}
where $b_{i}$ are some irrelevant constants and
\begin{equation}
a_i^{(1)}=4a_i^{(1/2)}=12a_i^{(0)}=\frac{1}{120\,(4\pi)^2}(-2,3)\,,
\end{equation}
with the superscript on the parameter $a_i^{(s)}$ referring to
vectors~$s=1$, fermions~$s=1/2$, and scalars~$s=0$. The
stress-energy tensor $\Tdw_{\mu\nu}$ is
\begin{displaymath}
T_{\mu\nu}^{\,\mathrm{DW}}=
    -\left(\begin{array}{cc}
        \eta_{\alpha\beta} & 0 \\
        0     & 0
    \end{array}\right)\sigma\,\delta(z)
\quad\Rightarrow\quad \hat{T}_{\mu\nu}^{\,\mathrm{DW}}=
    -\left(\begin{array}{cc}
        \eta_{\alpha\beta} & 0 \\
        0     & 0
    \end{array}\right)\sigma\,\delta^{\parallel}(p)\,,
\end{displaymath}
where $\parallel$ represents the directions along the DW, and
$\alpha\beta\dots$ are indices along these directions. With this,
the 1-loop contribution becomes, upon Fourier-transforming back to
space coordinates,
\begin{eqnarray}
h_{\alpha\beta}^{(1-loop)}&=&\kappa_1\frac{G_\mathrm{N}^2\sigma}{z}\eta_{\alpha\beta}\,,\\
h_{zz}^{(1-loop)}&=&-\kappa_2\frac{G_\mathrm{N}^2\sigma}{z}\, .
\end{eqnarray}
The parameters $\kappa_i$ are found to be $\kappa_1=0$ as the
contributions from $\Pi_1$ and $\Pi_2$ cancel, and
\begin{eqnarray}
\kappa_2=\frac{a}{120\pi}=-\frac{N^2}{4\pi}\,,
\end{eqnarray}
where $a=\sum_sN^{(s)}a_1^{(s)}=-5N^2$ for the field content of
$\mathcal{N}=4$ Super Yang-Mills theory with gauge group $SU(N)$.
Hence, the correction to the linearized metric due to the 1-loop
diagrams of the CFT has $h_{\alpha\beta}^{(1-loop)}=0$ and
$$
h_{zz}^{(1-loop)}=\frac{G_\mathrm{N}^2N^2}{4\pi\,z}\sigma~.
$$
This is clearly of pure gauge form, so we conclude that the CFT
does not correct the field created by a DW at 1-loop. Let us
insist that this is no longer true once the trace anomaly is
properly accounted for, which is possible only if we make an
ansatz where the DW is already inflating.

Let us end by noting that the result of this Appendix is expected
to hold at all orders in the loop expansion. This can be proven in
an elegant way by performing a K\"allen-Lehmann decomposition of
the graviton propagator. This is going to have a pole at $p^2=0$
and a branch cut, which is dual to the massless graviton and the
continuum of massive Kaluza Klein modes that appear in RS. Given
that the massive gravitons couple to matter through the
combination $h_{\mu\nu}\sim \Delta(p)
[T_{\mu\nu}-(T/3)\eta_{\mu\nu}]$, for relativistic Domain Walls
this always leads to a pure gauge form. Hence, massive gravitons
do not couple to the walls at least to linear order, and
irrespective of the actual form of the form factor $\Delta$ or
$\Pi_i$, the CFT radiative corrections at any loop order should
vanish.

\bibliographystyle{utphys}
\bibliography{dwRS}

\end{document}